\documentclass[onecolumn,epjc3]{svjour3}
\RequirePackage{fix-cm}
\smartqed
\RequirePackage{graphicx}
\RequirePackage{subfig}
\RequirePackage{float}
\usepackage{rotating}
\usepackage{amsmath,amssymb}
\usepackage{mathtools}
\RequirePackage[colorlinks,citecolor=blue,urlcolor=blue,linkcolor=blue]{hyperref}
\usepackage{doi}
\usepackage{siunitx}
\usepackage{xspace}
\usepackage[switch]{lineno}
\usepackage{cite}
\usepackage{comment}
\usepackage{url}
\usepackage[version=4]{mhchem}
\usepackage{upgreek}
\usepackage{booktabs}
\usepackage{bold-extra}

\newcommand{\cupid}{CUPID}

\newcommand{\DBD}{$0\nu\beta\beta$}

\newcommand{\Mo}{\ce{^{100}Mo}}

\newcommand{\LMO}{\ce{Li$_2$MoO$_4$}}
\newcommand{\lmo}{LMO}

\newcommand{\BDPT}{GDPT}
\newcommand{\ckky}{cts$/($keV$\cdot$kg$\cdot$yr$)$}

\journalname{Eur. Phys. J. C}
\begin{document}


\title{Innovating Bolometers' Mounting: A Gravity-Based Approach}

\author{K.~Alfonso\thanksref{VT_US,UCLA_US}
\and A.~Armatol\thanksref{LBNL_US,fn0}
\and C.~Augier\thanksref{IP2I_France}
\and F.~T.~Avignone~III\thanksref{UofSC_US}
\and O.~Azzolini\thanksref{LNL_Italy}
\and A.S.~Barabash\thanksref{NRC_KI_Russia}
\and G.~Bari\thanksref{SdB_Italy}
\and A.~Barresi\thanksref{UniMIB_Italy,MIB_Italy}
\and D.~Baudin\thanksref{CEA_IRFU_France}
\and F.~Bellini\thanksref{SdR_Italy,SURome_Italy}
% ...
}

\institute{Virginia Polytechnic Institute and State University, Blacksburg, VA, USA\label{VT_US}
\and
University of California, Los Angeles, CA, USA\label{UCLA_US}
\and
Lawrence Berkeley National Laboratory, Berkeley, CA, USA\label{LBNL_US}
\and
Univ Lyon, Universit\'e Lyon 1, CNRS/IN2P3, IP2I-Lyon, Villeurbanne, France\label{IP2I_France}
\and
University of South Carolina, Columbia, SC, USA\label{UofSC_US}
\and
INFN Laboratori Nazionali di Legnaro, Legnaro, Italy\label{LNL_Italy}
\and
National Research Centre Kurchatov Institute, Kurchatov Complex of Theoretical and Experimental Physics, Moscow, Russia\label{NRC_KI_Russia}
\and
INFN Sezione di Bologna, Bologna, Italy\label{SdB_Italy}
\and
Universit\`a degli Studi di Milano-Bicocca, Dipartimento di Fisica, Milano, Italy\label{UniMIB_Italy}
\and
INFN Sezione di Milano-Bicocca, Milano, Italy\label{MIB_Italy}
\and
IRFU, CEA, Universit\'e Paris-Saclay, Saclay, France\label{CEA_IRFU_France}
\and
INFN Sezione di Roma, Rome, Italy\label{SdR_Italy}
\and
Sapienza University of Rome, Rome, Italy\label{SURome_Italy}
% ...
\and
INFN Laboratori Nazionali del Gran Sasso, Assergi (AQ), Italy\label{LNGS_Italy}
\and
INFN Laboratori Nazionali di Frascati, Frascati, Italy\label{LNF_Italy}
\and
CNR-Institute for Microelectronics and Microsystems, Bologna, Italy\label{CNR-IMM_Italy}
\and
INFN Sezione di Pavia, Pavia, Italy\label{PV_Italy}
% ...
\and
University of California, Berkeley, Berkeley, CA, USA\label{UCB_US}
\and
Universit\'e Paris-Saclay, CNRS/IN2P3, IJCLab, Orsay, France\label{IJCLab_France}
\and
INFN Sezione di Genova, Genova, Italy\label{SdG_Italy}
% ...
\and
Rome Tor Vergata University, Rome, Italy\label{Rome_Tor_Vergata_University_Italy}
\and
INFN sezione di Roma Tor Vergata, Rome, Italy\label{INFN_Tor_Vergata_Italy}
\and
Northwestern University, Evanston, IL, USA\label{NWU_US}
\and
Argonne National Laboratory, Argonne, IL, USA\label{ANL_US}
% ...
\and
Massachusetts Institute of Technology, Cambridge, MA, USA\label{MIT_US}
\and
Boston University, Boston, MA, USA\label{BU_US}
\and
California Polytechnic State University, San Luis Obispo, CA, USA\label{CalPoly_US}
% ...
\and
Yale University, New Haven, CT, USA\label{Yale_US}
\and
Department of Physics and Astronomy, University of Pittsburgh, Pittsburgh, PA, USA\label{Pittsburgh_US}
\and
Drexel University, Philadelphia, PA, USA\label{Drexel_US}
\and
Johns Hopkins University, Baltimore, MD, USA\label{JHU_US}
% ...
\and
Nikolaev Institute of Inorganic Chemistry, Novosibirsk, Russia\label{NIIC_Russia}
\and
INFN Sezione di Padova, Padova, Italy\label{PD_Italy}
\and
Univ. Grenoble Alpes, CNRS, Grenoble INP, SIMAP, Grenoble, France\label{SIMaP_Grenoble_France}
\and 
University of Bologna, Bologna, Italy\label{UnivBologna_Italy}
}

\thankstext{fn0}{Now at IP2I-Lyon, University of Lyon, France}
\thankstext{fn1}{Now at Physik-Institut, University of Z\"urich, Z\"urich, Switzerland}
\thankstext{fn2}{Now INFN Laboratori Nazionali di Frascati, Frascati, Italy}
\thankstext{fn3}{Also at Instituto de F{\'\i}sica, Universidade de S\~ao Paulo, Brazil and Max-Planck-Institut f\"ur Physik, M\"unchen, Germany}


\date{Received: date / Accepted: date}

\maketitle

\begin{abstract}
Cryogenic calorimeters, also known as bo\-lo\-me\-ters, are among the leading technologies for searching for rare events. The CUPID experiment is exploiting this technology to deploy a tonne-scale detector to search for neutrinoless double-beta decay of $^{100}$Mo.
The CUPID collaboration proposed an innovative approach to assembling bolometers in a stacked configuration, held in position solely by gravity. This gravity-based assembly method is unprecedented in the field of bolometers and offers several advantages, including relaxed mechanical tolerances and simplified construction. To assess and optimize its performance, we constructed a medium-scale prototype hosting 28 \LMO\ crystals and 30 Ge light detectors, both operated as cryogenic calorimeters at the Laboratori Nazionali del Gran Sasso (Italy). Despite an unexpected excess of noise in the light detectors, the results of this test proved (i) a thermal stability better than $\pm$0.5\,mK at 10 mK, (ii) a good energy resolution of \LMO\ bolometers, (6.6 $\pm$ 2.2)\,keV FWHM at 2615\,keV, and (iii) a \LMO\ light yield measured by the closest light detector of 0.36\,keV/MeV, sufficient to guarantee the particle identification re\-ques\-ted by CUPID. 

\end{abstract}



\section{Introduction}
Discovering the existence of hypothetical neutral massive fermions, such as Majorana neutrinos~\cite{Majorana1973}, would be a significant breakthrough in particle physics. 
All phenomena observed to date conserve the total lepton number~\cite{Navas2024}; however, the existence of Majorana fermions would introduce processes where this number is violated~\cite{RevModPhys.95.025002}. Among these processes, neutrino-less double-beta decay (\DBD) is especially intriguing as it would result in the creation of two electrons, meaning two matter particles, and no anti-neutrinos~\cite{Furry}. The profound implications of such a discovery across various sectors of physics~\cite{RevModPhys.95.025002} have driven an extensive experimental efforts dedicated to the search for \DBD.

The observable of this process is the half-life of the few nuclei for which \DBD\ is energetically possible ($T_{1/2}^{0\nu}$). Today, the most competitive experiments are setting lower limits on $T_{1/2}^{0\nu}$ on the level of 10$^{26}$\,yr~\cite{Anton2019,Agostini:2020,Abe2024,adams2024,Arnquist2023}.
In the hypothesis that the \DBD\ is mediated through light Majorana neutrino exchange~\cite{RevModPhys.95.025002}, 
$T_{1/2}^{0\nu}$ can be expressed as a function of the effective Majorana neutrino mass m$_{\beta\beta}$.
The current generation of experiments is reaching and entering the region of m$_{\beta\beta}$ corresponding to the inverted hierarchy of the neutrino masses.
Next-generation projects aim at improving the sensitivity on $T_{1/2}^{0\nu}$ by at least one order of magnitude. This will allow to reach a complete coverage of the inverted hierarchy region -- which also corresponds to about 50$\%$ of the most probable region of the normal hierarchy~\cite{Agostini:2017jim}, with the hope of discovering the \DBD\ signal.

CUPID (CUORE Upgrade with Particle IDentification~\cite{CUPID2024_bsl}) is a next-generation experiment based on the technique of cryogenic calorimeters, historically also called ``bolometers"~\cite{Fiorini:1984}. In bolometers, thermal variations caused by energy deposits are measured with exquisite sensitivity using thermal sensors. 
CUORE~\cite{ALDUINO20199,Alduino:2018,Adams:2022}, a ton-scale detector currently running at the Laboratori Nazionali del Gran Sasso (LNGS) in Italy, is the most sensitive \DBD\ search experiment based on this technology.

In contrast to its predecessor CUORE, CUPID will deploy bolometers with capability of reading scintillation light, enabling particle identification~\cite{Pirro:2005ar,poda2021}. 
This feature is critical to suppress the dominant CUORE background, i.e., $\alpha$ particles produced by naturally radioactive contaminants of the surface of the detector mechanical structure~\cite{Adams.PhysRevD.110.052003}.
To this aim, CUPID will deploy \LMO\ (LMO) crystals, which light output is about one order of magnitude larger compared to the TeO$_2$ crystals of CUORE \cite{poda2021}.
A second important upgrade compared to CUORE consists in the use of a high Q-value emitter, $^{100}$Mo. The Q-value of $^{100}$Mo, (3034.40  $\pm$ 0.17) keV~\cite{Rahaman:2008}, lies in a region where the natural $\gamma/\beta$ radioactivity is suppressed by more than 10 times according to the data-driven CUORE background model \cite{CUORE:2024fak}.

The dual read-out of heat and light, as well as the potential of using high Q-value emitters, were already demonstrated by two pilot experiments\footnote{Small-scale experiments based on molybdate scintillating bolometers, AMoRE-pilot~\cite{Alenkov2019} and AMoRE-I~\cite{agrawal2024background}, were run at the Yangyang underground laboratory (Korea), setting the most stringent limit on the \DBD\ of $^{100}$Mo~\cite{agrawal2024}. These experiments are stages of the AMoRE project~\cite{alenkov2015}, a large-scale experiment which is not part of the CUPID program and uses a different technology for thermal sensing (metallic magnetic calorimeters).}, CUPID-0~\cite{Azzolini:2018tum,Azzolini-cupid0-final} and CUPID-Mo~\cite{Armengaud:2019loe,Augier:2022znx}. The former consisted of two natural and 24 ZnSe crystals $\sim$95$\%$ enriched in $^{82}$Se, for a total active mass of 10.5\,kg, and was operated between 2017--2020 at the LNGS. CUPID-Mo was made of 20 \LMO\ crystals $\sim$97$\%$  enriched in $^{100}$Mo with a mass of $\sim$0.2\,kg each, and it was operated for 1.5\,yr (2019--2020) at the Laboratoire Souterrain de Modane in France.
Both CUPID-0 and CUPID-Mo proved a rejection of $\alpha$ particles above 99.9\,\%, reaching a background in the region of interest in the scale of 10$^{-3}\,$\ckky~\cite{Azzolini2018b,Azzolini:2019nmi,Augier2023-bkg-model}. Furthermore, despite the small exposure, they were able to set competitive limits on many physics processes~\cite{Azzolini:2018oph,Azzolini:2018dyb,Azzolini:2019tta,Azzolini_2019,Azzolini2020:Zn,Azzolini_2019-lorentz,Azzolini_2021-216po,Azzolini_2023,azzolini2023-2nu,Augier2023-ExcitedStates,Armengaud_2021,CUPID-Mo:2023lru,augier2024}, proving the potential of this technology.

Building on the results of its demonstrators, the \cupid\ detector will consist of a close-packed array of 1596 LMO bolometers $\sim$95$\%$ enriched in $^{100}$Mo. The LMO bolometers (45 $\times$ 45 $\times$ 45\,mm$^3$ each, for a total mass of 240\,kg of $^{100}$Mo) will be interleaved by bolometric light detectors (LDs)~\cite{CUPID2024_bsl}.
The devices will be divided in 57 towers, each of them consisting of 14 floors with two modules per floor.

In all the previous bolometric experiments to search for $0\nu\beta\beta$, the crystals were kept in position using polytetrafluoroethylene (PTFE)  pieces inserted in a rigid copper structure consisting of columns and frames (see, for example, ~\cite{Alduino2016b,Azzolini:2018tum,Armengaud:2019loe,Kim:2022uce}). 
However, this approach has several limitations, such as strict mechanical tolerances, complicated assembly, and challenging cleaning due to sharp-edged components and to the presence of threads.
In this study, we explore a novel solution to address these issues. Our approach involves a fully floating mechanical structure: each floor is stacked atop the previous one using only gravity, without columns and screws.
Following the results obtained with a small-scale prototype~\cite{Alfonso2022}, based on 8 CUPID-size LMOs, we developed and assembled an array with the same size of a CUPID tower, called the Gravity Design Prototype Tower (\BDPT).
In the following sections, we detail the design, construction process, operation at low tem\-pe\-ra\-tu\-re, and performance of the detector.

\section{Gravity Design Prototype Tower}
\begin{figure*}[!thbp]
  \centering
   \includegraphics[height=0.3\textheight]{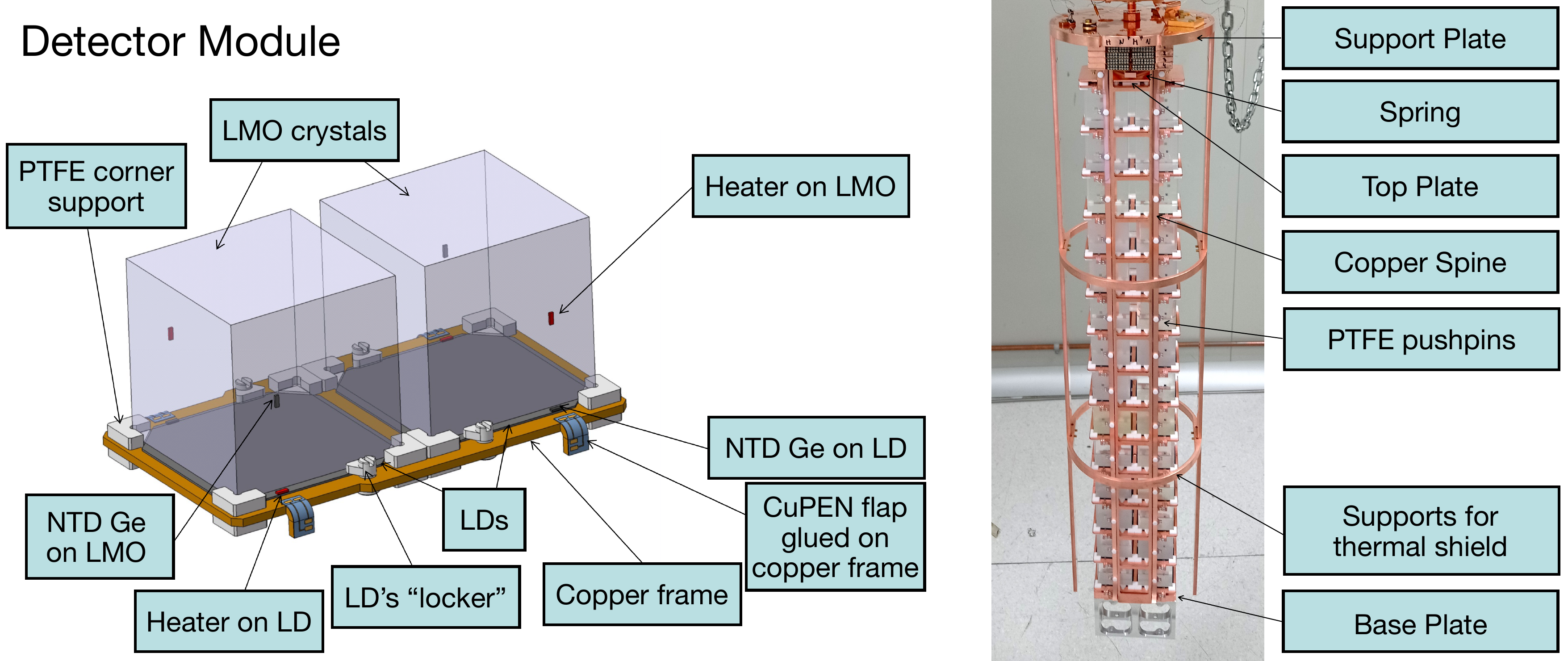}
  \caption{Left: single module of the CUPID tower~\cite{Alfonso2022}. A copper frame is equipped with PTFE ``corners" to guide the positioning of the octagonal LD (back squares); the LDs are then locked using two PTFE elements. Two LMO crystals are positioned on the PTFE ``corners" at a distance of 0.5\,mm from the LDs. Right: tower  with 28 LMO bolometers and 30 cryogenic Ge LDs appended to the coldest point of the refrigerator.}
  \label{fig:CUPID-detector}
\end{figure*}
The design of the first CUPID tower prototype was driven by the following requirements: reducing constraints on mechanical tolerances, increasing the filling factor of the pre-existing cryostat, aiming for a small amount of inert material (copper/LMO $<$ 20\%), minimizing machining of components to reduce potential radioactive contamination, and ensuring ease in the production, cleaning, and construction of the detector.

The first version of this design, the \BDPT, consists of 14 floors of LMOs interleaved by LDs (30 in total, as each \lmo\ is equipped with LDs both on top and bottom).
Each module includes two \lmo s and their bottom LDs, as depicted in Fig.~\ref{fig:CUPID-detector}-left.

A 2\,mm thick copper frame, produced by laser cutting, is fitted with 8 PTFE ``corners" (four elements for each \lmo). The octagonal LD sits atop the frame and is secured in place using two PTFE ``lockers". The two \lmo s are positioned on the PTFE ``corners". 

The copper holders and PTFE elements were designed to ensure that the LDs are positioned as close as possible to the LMO crystals (0.5\,mm, in contrast to the design of the predecessors that allowed for a minimum distance of 4\,mm~\cite{Armatol:2021}), maximizing light collection efficiency.
In contrast to CUPID-Mo, we opted to eliminate the reflecting foil around the \lmo\ crystals. Initial studies on prototypes with a small number of CUPID-size crystals indicated that the reflector is unnecessary for achieving adequate light collection~\cite{Alfonso2023,Alfonso2022}. Moreover, its presence would hinder the complete identification of background events that do not entirely occur within a single crystal. Removing the reflector also reduces the mass of passive components, simplifies detector construction and modeling.

The modules are stacked into a tower reaching a height of 14 floors (Fig.~\ref{fig:CUPID-detector},right). The topmost \lmo\ floor is enclosed by a copper frame that houses the final two LDs. Two copper plates, one at the top (``top plate") and one at the bottom (``base plate") complete the tower.
On the sides of the detector, two copper spines are anchored exclusively to the top and base plates. These spines support the entire weight of the tower and serve as pathways for the electrical connections (Section~\ref{sub:CuPEN}).
Additionally, a copper structure moun\-ted on the top plate supports a spring capable of applying up to 12 kg of pressure to the entire array. This value, identified as optimal through engineering simulations, ensures the stability of the array while preventing any risk of damage.

The copper spines are designed to have no mechanical connection with the frames, which are kept in position solely by gravity. However, due to a design oversight, the spines experienced some friction with the frames, which could potentially induce noise in the detector.

While the LMO sensors are directly wire-bonded to the copper strips along the spines, the LD sensors require an intermediate step. Initially, they are pre-bonded to copper strips on the copper frames. After assembling the tower, a second wire-bonding connects these copper strips to those on the vertical spines.
This electrical connection also ensures the thermalization of each frame.

In the following, we describe more in detail the tower composition.

\subsection{GDPT \lmo\ Diversity}
\label{sec:LMO}
The \BDPT\ was constructed using \lmo\ crystals already owned by participating institutions in the project. Most of them had been previously tested in various R$\&$D activities. 
The LMOs are cubes with a side length of 45\,mm and consist of:
\begin{itemize}
\item 14 natural \lmo\ crystals produced at the  Nikolaev Institute of Inorganic Chemistry (NIIC, Novosibirsk, Russia) by low-thermal\-gradient Czochralski technique following the LUMINEU protocol on purification and crystalization~\cite{Berge:2014,Armengaud:2017hit,Grigorieva:2017}, a\-dop\-ted for CUPID-Mo as well~\cite{Armengaud:2019loe}; 6 out of these 14 crystals were provided by the BINGO project~\cite{Armatol:2024bsb}, while the rest samples were produced for CUPID R$\&$D activities~\cite{Alfonso2022};
\item 8 \lmo\ crystals 98$\%$ enriched in $^{100}$Mo and characterized within the CROSS project~\cite{Armatol_2021,Alfonso2023}; these crystals (not ideal cubes, as they have chamfered edges) were also produced at the NIIC by low-thermal-gradient Czochralski crystal growth following the LUMINEU protocol;
\item 6 natural \lmo\ crystals produced in China as a first step of an iterative process (growth - test) aiming at developing a growth protocol; 
2 of them grown by Bridgman method at the Shanghai Institute of Ceramics, Chinese Academy of Sciences (SICCAS) and
4 of them grown at the Ningbo University at Zhejiang by Czochralski method.
\end{itemize}

\begin{figure}[htbp]
  \centering
  \includegraphics[height=0.7\textwidth]{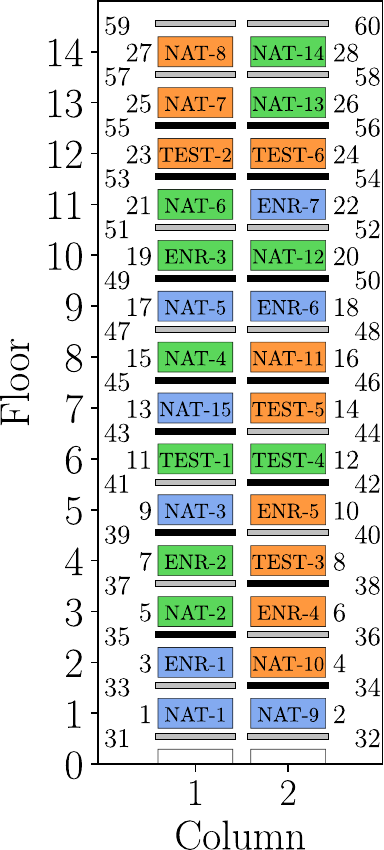}
  \caption{Schematic view of the tower composition. Big rectangles represent LMOs. Crystals produced at the NIIC are labeled as ENR-1, \dots, ENR-7 (enriched crystals) and NAT-1, \dots, NAT-15 (natural crystals). TEST-1, \dots, TEST-6 refer to the natural crystals grown at the Ningbo University and at the SICCAS. Different colors refer to different glue types: blue (UV-polymerization glue), orange (A\-ral\-di\-te\textsuperscript{\textregistered} ~Rapid), green (A\-ral\-di\-te\textsuperscript{\textregistered} ~Slow). LDs are depicted with thin rectangles whose color indicates the anti-reflective deposition technique: black (evaporation) and gray (sputtering). Numbers close to the LMOs (1--28) and to the LDs (31--60) indicate the channel number.}
  \label{fig:detector-scheme}
\end{figure}

The position of each crystal in the tower is illustrated in Fig.~\ref{fig:detector-scheme}.
Each LMO crystal is equipped with a thermal sensor consisting of a 3 $\times$ 3 $\times$ 1\,mm$^3$ Neutron Transmutated Doped (NTD) Ge thermistor~\cite{Haller}. This device converts the temperature variation of the LMOs into a readable voltage signal.  

In previous experiments (including CUORE, CUPID-0 and CUPID-Mo), NTD Ge thermistors were attached to the crystals using the bi-component  epoxy glue A\-ral\-di\-te\textsuperscript{\textregistered} ~Rapid, chosen for its excellent performance at cryogenic temperatures and high radiopurity. However, the fast curing time of A\-ral\-di\-te\textsuperscript{\textregistered} Rapid ($\approx$ 5 minutes), limited the gluing operation to one crystal at a time in past experiments.
Switching to A\-ral\-di\-te\textsuperscript{\textregistered} Slow, with a curing time of about 1 hour, would allow multiple crystals to be glued in a single batch, significantly accelerating the mass production process for CUPID. Therefore, the \BDPT\ was used for the first high-statistics test of this glue.

Preliminary studies on the UV-polymerization glue PERMABOND\textsuperscript{\textregistered} 620 indicated it as a promising alternative to A\-ral\-di\-te\textsuperscript{\textregistered} epoxy~\cite{Auguste:2024xrg}. The potential to deposit much thinner layers could, in theory, reduce the interface between the crystal and the sensor, resulting in faster pulses. Moreover, the curing time is very fast, tens of seconds under an UV lamp. The successful results from R$\&$D with this glue prompted us to use this method for part of the \BDPT\ crystals.
LMO bolometers with different types of glue are placed in various positions within the tower to ensure that any observed differences can be attributed solely to the glue type (Fig.~\ref{fig:detector-scheme}).

Finally, each LMO is equipped with a Si heater for thermal gain stability control~\cite{Alfonso:2018a:pulser,Alessandrello:1998bf,Andreotti:2012zz}.
\subsection{GDPT Light Detectors}
The easiest way to operate a photon detector at low temperatures is to use a cryogenic calorimeter: impinging photons are converted into phonons and measured using thermal sensors. The LDs for this prototype are fabricated from 500\,$\upmu$m-thick high-purity Ge wafers cut into an octagonal shape to fit the detector structure. Each LD is coated with a 60\,nm-thick SiO anti-reflecting layer to increase light absorption~\cite{Mancuso:2014paa}. The anti-reflecting layer is deposited via evaporation on 14 of the LDs and via sputtering on the remaining 16. While in CUPID-0 and CUPID-Mo experiments used SiO deposited by evaporation, sputtering would be preferred for the future mass production of thousands of LDs. Therefore, the \BDPT\ was chosen as a benchmark to validate this deposition method.
To ensure that any observed differences in light collection can be attributed to the deposition technique, LDs with sputtered and evaporated SiO are alternated in the tower (Fig.~\ref{fig:detector-scheme}).

The LDs are equipped with a Si heater, identical to the one used for the LMOs, and with a 3 $\times$ 0.5 $\times$~1\,mm$^3$ NTD Ge thermistor. A smaller temperature sensor compared to the one of the LMOs ensures a smaller heat capacity, which is crucial given the tiny mass of the LDs.
Both the heaters and thermistors were attached using A\-ral\-di\-te\textsuperscript{\textregistered} epoxy glue.

\begin{figure*}[htbp]
  \centering
  \includegraphics[height=0.3\textheight]{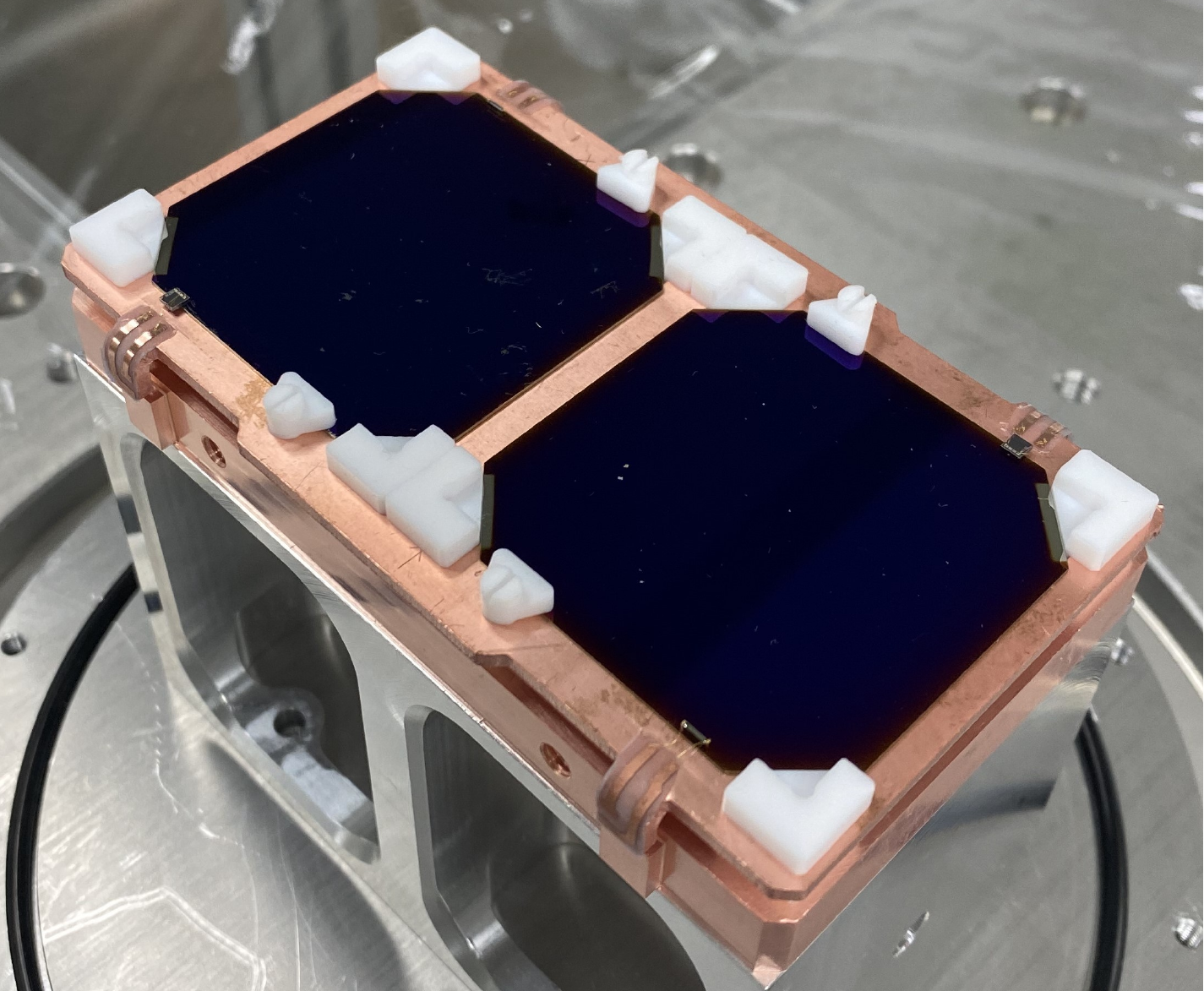}
    \includegraphics[height=0.3\textheight]{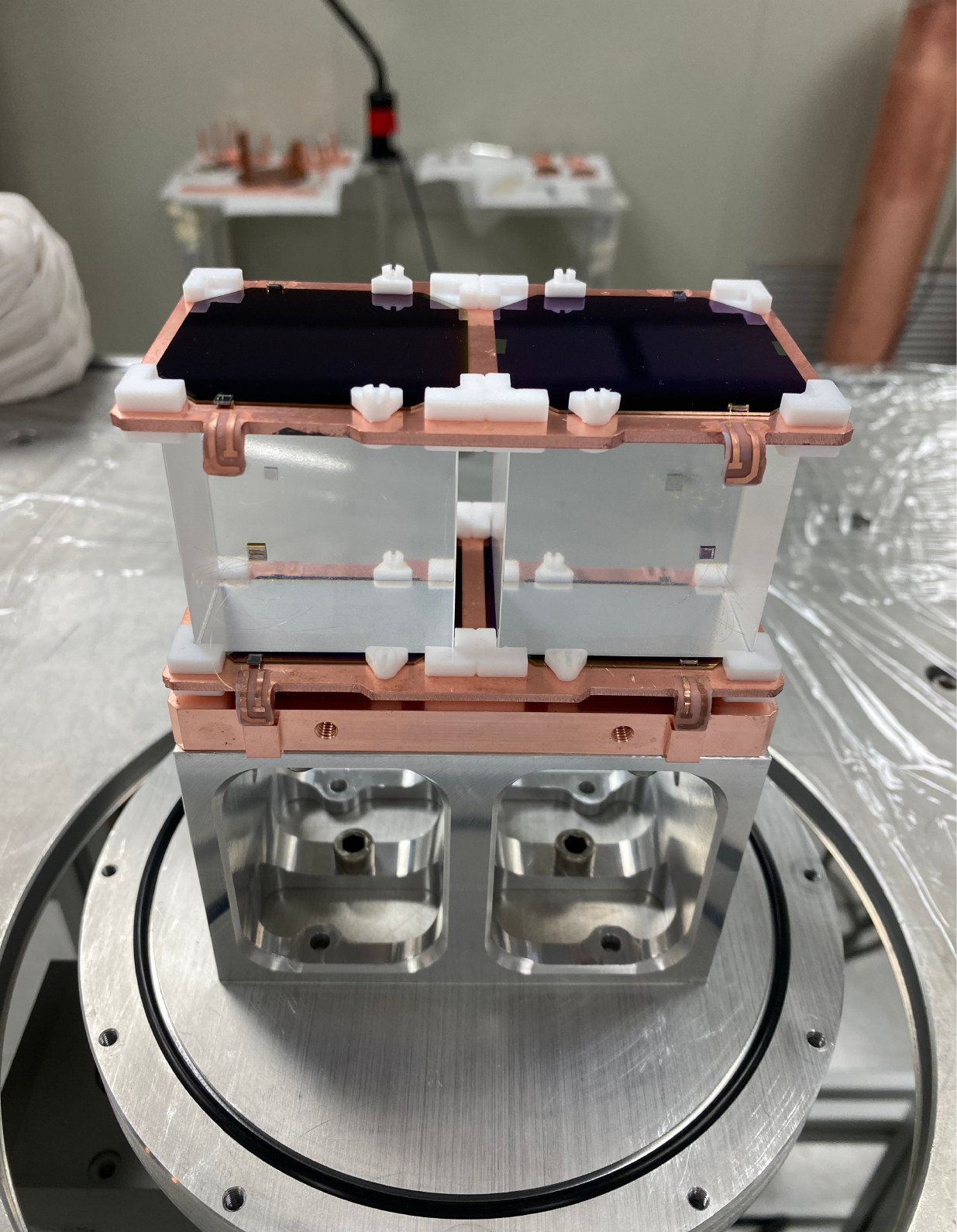}
  \caption{Left: two Ge light detectors are positioned on the copper frame using four PTFE ``corners" as guide, and secured using two PTFE ``lockers". This first frame is positioned on the copper ``base" plate. The electrical connections of the 3 $\times$ 0.5 $\times$ 1\,mm$^3$ NTD Ge thermistor and of the 2.4 $\times$ 2.4 $\times$ 0.5\,mm$^3$ silicon heater are made using a gold wire that was previously bonded on the pads of the sensors. The gold wire is then bonded to the CuPEN glued on the frames. Right: two LMO crystals are positioned on top of the LDs. Another floor of LDs is then placed atop the LMOs.
  This is the first module of the \BDPT. After mounting the whole tower, a second bonding is done between the CuPEN located on the frames and the CuPEN attached to the copper spines.}
  \label{fig:BDPT-LD-photo}
\end{figure*}

The LDs are placed on the copper frames by using the PTFE ``corners" to guide their positioning. They are then secured using the two PTFE ``lockers", as shown in Fig.~\ref{fig:BDPT-LD-photo}.
Their NTD Ge thermistors and Si heaters were wire-bonded to the copper strip located on the frame itself.

Before installing these devices in the \BDPT, we performed a test with four of them in a pulse-tube cryostat at the surface laboratory of IJCLab (Orsay, France), where the noise environment induced by the pulse-tube cryocooler is not mitigated as in the CUORE cryostat. The devices were only spring-suspended inside the cryostat to reduce the noise induced by vibrations~\cite{Olivieri:2017lqz}.
Despite the sub-optimal noise conditions, all the LDs achieved performance in compliance with the CUPID requirements: a noise RMS resolution between 70 and 90\,eV~\cite{Alfonso_2023-pulsetube}, compared to the CUPID target of $<$100\,eV~\cite{CUPID2024_bsl}.

\subsection{Cryogenic Readout: CuPEN}
\label{sub:CuPEN}
The cryogenic readout cabling is made of CuPEN (Copper on a Polyethylene 2.6 Naphthalate substrate) tapes. The design of the CuPEN in the \BDPT\ is adapted from the one already tested within the CUORE experiment~\cite{Andreotti:2009zza} to include the readout of the LDs. The heaters of the LDs are wired on the same tape as the NTD Ge thermistors of the LMOs and vice versa, to minimize cross-talk.

Before assembling the tower, the CuPEN was glued onto the copper spines using A\-ral\-di\-te\textsuperscript{\textregistered} glue. To help keep the CuPEN in position, small PTFE pushpins were mounted on the copper spines.

In CUPID, the CuPEN will be plugged into Zero Insertion Force (ZIF) connectors mounted on Kapton\textsuperscript{\textregistered} boards at the coldest point of the dilution refrigerator (mixing chamber plate). However, for the \BDPT\ we opted for a custom-made connector assembled on top of the tower to utilize the pre-existing readout of the cryostat used for the test.

\subsection{Assembly}
\label{sec:construction}
The procurement of all the components for the \BDPT, as well as the construction of its assembly station, began in mid-2021. The construction of the \BDPT\ was completed in May 2022. 
The ``Assembly Line" was conceived and designed keeping in mind the need to construct the CUPID detector in a controlled and safe atmosphere, which will be ensured by application of dedicated and ergonomic glove boxes per each assembly step. 

For the \BDPT\ assembly, a protective volume capable of maintaining the detector in nitrogen atmosphere was planned only for the storage phase, which poses the highest risk of re-contamination due to exposure times. 
The detector construction was carried out with the help of a motorized and automated machinery capable of moving the detector along the vertical axis and rotating it around its central axis.

\begin{figure}[htbp]
  \centering
    \includegraphics[height=0.37\textheight]{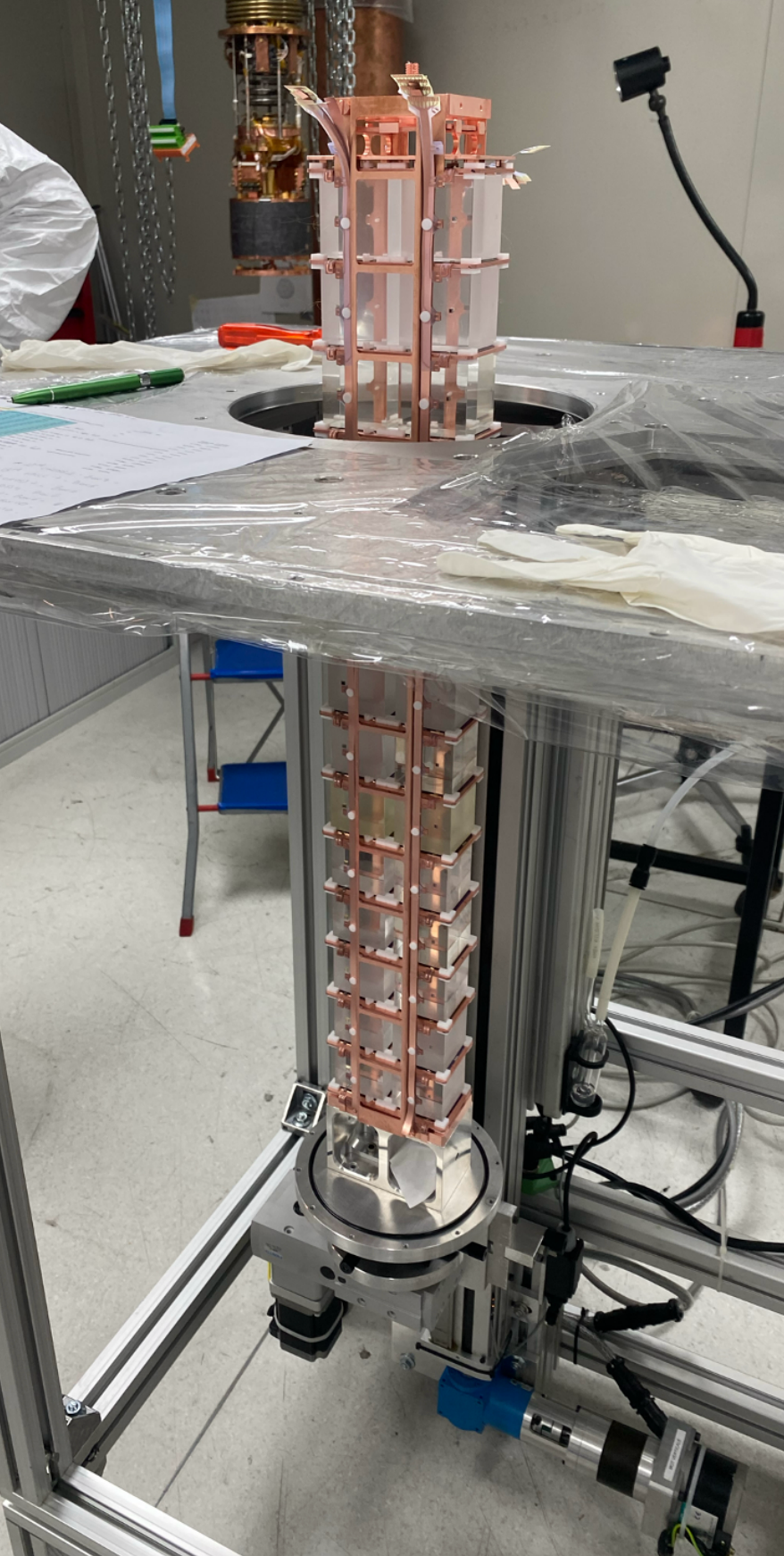}
  \includegraphics[height=0.37\textheight,trim={0 0 0cm 0},clip]{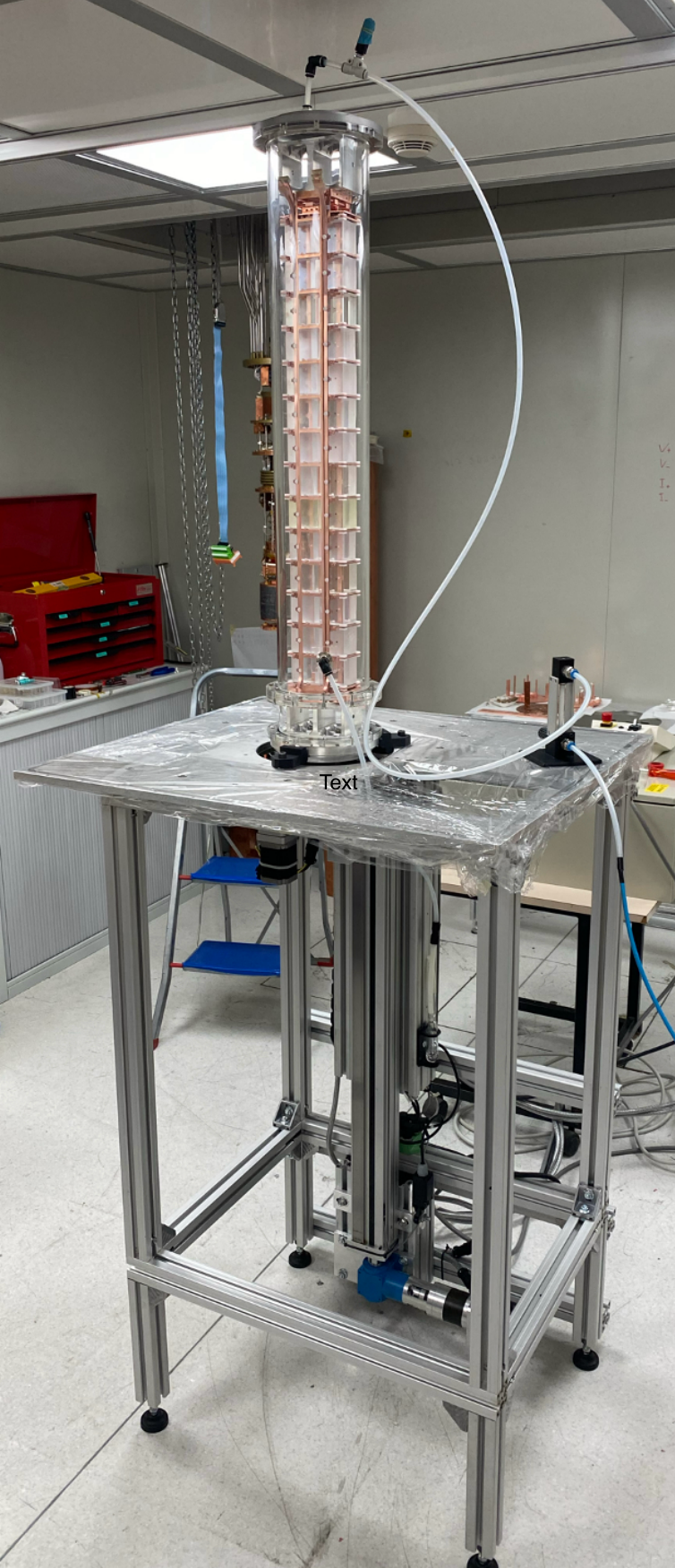}  
     \caption{Tooling for the detector tower assembly. Left: an automated lifting and rotation system allows for a convenient work height during the tower assembly. Right: the circular tower base plate interfaces with an acrylic capsule for transport and storage in an inert atmosphere.}
  \label{fig:CUPID-tower}
\end{figure}

This tooling, called ``Garage" in relation to its function of protecting the detector under construction, enabled to proceed with a rapid and smooth assembly of all 14 floors within one day. 
The ``Garage" (Fig.~\ref{fig:CUPID-tower}-left) was designed, built and tested at INFN-Roma, performing an upgrade of the machinery already existing and used for the assembly of the CUPID-0 detector~\cite{Azzolini:2018tum}. 
The strokes of the various automations were adapted to the geometry of the tower, which is larger in all dimensions. The translation and rotation speeds were optimized to ensure appropriate sensitivity in the planned manual operations of positioning the crystals and frames with the LDs already assembled and pre-bonded, while also keeping the ergonomics of the operations as a criterion.

The ``Garage" is designed to be able to be sealed with a protective capsule. This capsule forms a secure container which allows the tower to be lifted, moved and stored under continuous flow of nitrogen in order to preserve its cleanliness and avoid re-contamination. 
Such a capsule -- named ``Storage Box" (Fig.~\ref{fig:CUPID-tower}-right) -- is a chamber composed of transparent acrylic plastic, so that the detector can be inspected at any time, and anodized aluminum alloy flanges to ensure sufficient surface stability.
The ``Storage Box" was equipped with a gas flushing system capable of ensuring fine adjustment of the inlet flow and a constant internal overpressure of about 2\,mbar by means of an appropriate non-return valve.
The machine was commissioned at INFN-Roma and installed in the clean-room of the CUPID Hall A cryostat at the LNGS. 

The tower was assembled on May 4th, 2022, starting from LDs and LMOs already equipped with pre-wired NTDs and heaters. The gold wires attached to the sensors were soldered onto the CuPEN pads using indium. When constructing the \BDPT, the bonding machine and bonding station that will be used in CUPID were not yet available.

\section{Installation and Data Taking}

The GDPT was installed in an Oxford TL-1000 dilution refrigerator, equipped with a liquid He bath. 
The cryostat hosted the CUORE-0 experiment~\cite{Alduino2016b} and its damping system and electronics were later upgraded for the CUPID-0 demonstrator, as described in~\cite{Azzolini:2018tum}. 
The GDPT was hung from a copper plate and then connected to the cryostat's mixing chamber through a copper spring to reduce vibrations.

The custom-made connectors were attached to the pre-existing readout of the cryostat with the help of silk-covered Constantan\textsuperscript{\textregistered} ~twisted pairs. The silicon heaters for each column of the GDPT and each detector class (LMO crystals and LDs) were operated in parallel, totaling four parallel heater lines. Several NTD Ge thermometers have been installed along the tower to monitor its temperature uniformity. In particular, one thermometer was placed on the tower ``top'' support plate, another on the 7th floor, and a third one on the ``base'' support plate.

The data acquisition system consisted of an amplification stage ~\cite{Azzolini:2018tum, Arnaboldi:2017aek}, a six-pole anti-aliasing active Bessel filter with programmable cutoff frequency, and an 18-bit resolution ADC. The data stream was digitized at 1\,kHz for LMOs and 2\,kHz for LDs and stored on disk in NTuples using a ROOT-based software framework (APOLLO), developed for CUORE~\cite{DiDomizio:2018ldc}. 
An online software derivative trigger algorithm identified physical pul\-ses and randomly selects noise events. The derivative trigger threshold was optimized for each channel. 
A detailed description of the experimental setup, electronics, and DAQ can be found in~\cite{Azzolini:2018tum}.

\begin{figure}
    \centering
    \includegraphics[width=1\linewidth]{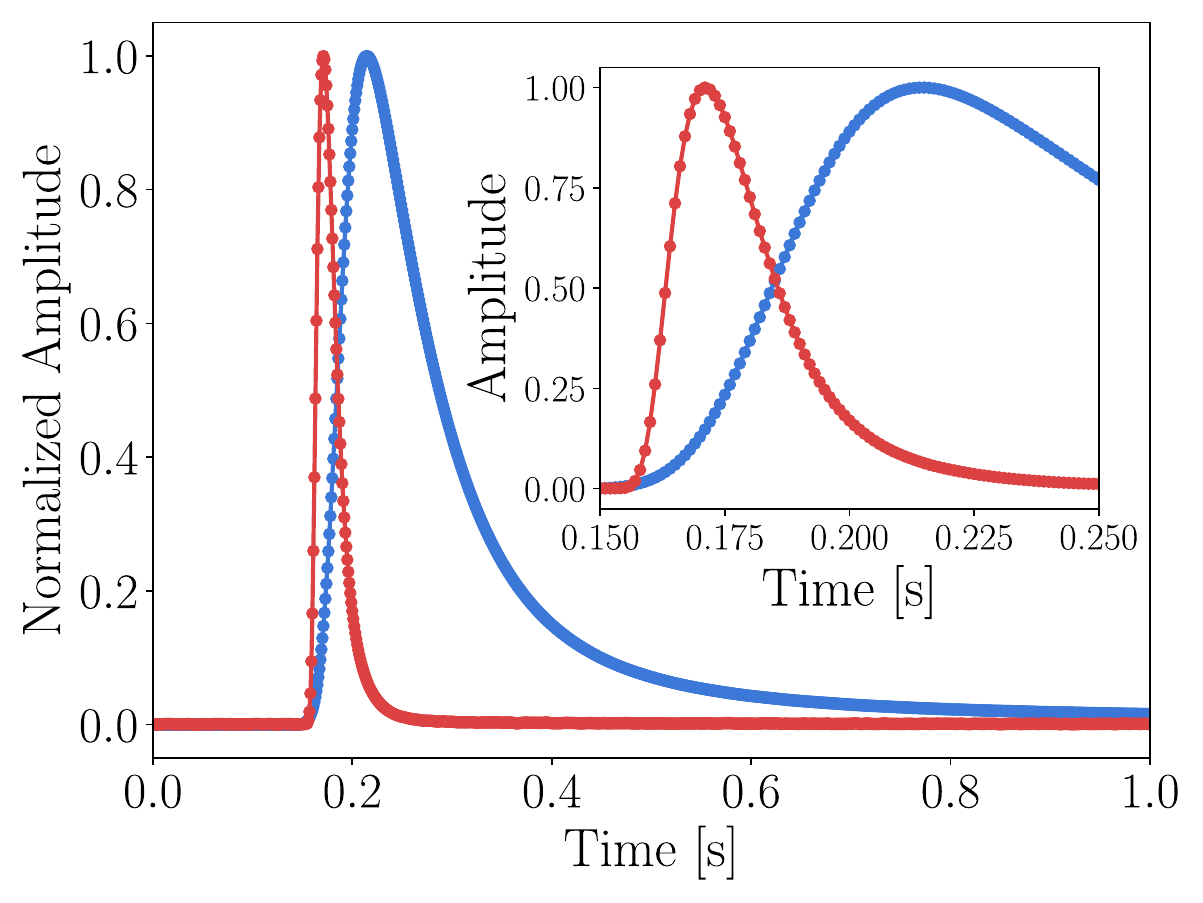}
    \caption{Average pulse templates for LMO and LD sample channels (LMO-1 and LD-31), depicted in blue and red, respectively. The amplitude of the two pulses has been normalized to one. The inset highlights the different time development of the two detectors.}
    \label{fig:AP}
\end{figure}

The analysis procedure utilized the DIANA signal processing framework~\cite{Alduino:2016} and followed an approach similar to that described in~\cite{Azzolini2018b,Armatol_2021}. 
Pre-trigger and post-trigger time intervals were chosen to contain the complete thermal pulse time development. Since heat and light pulses had different time responses, time windows for LMOs and LDs were set to 2\,s and 0.6\,s, respectively. In both cases the pre-trigger time was set to be half of the window.
From each triggered window, we calculated various parameters, including the number of triggers in the acquisition window, the slope of the baseline (average of the pre-trigger), and the rise time and decay time of the pulses. These parameters were then used to select typical signal and noise events, from which we obtained the average pulse (see Fig.~\ref{fig:AP}) and the average noise templates. These templates were crucial for applying the Optimum Filter~\cite{Gatti:1986}, a technique to estimate the amplitude of each triggered event by maximizing the signal-to-noise ratio.

\(^ {232}\)Th calibration sources with different intensities were placed outside the cryostat during data taking to reconstruct the energy response of our detectors. 
LMO bolometers were calibrated on the \(^ {208}\)Tl peak at 2615\,keV, while LDs were calibrated on the most prominent Mo X-rays peak (at approximately 17.4\,keV~\cite{thompson2001x}), whose emission was induced by a higher intensity \(^ {232}\)Th source, similar to the method employed in LUMINEU~\cite{Armengaud:2017hit} and CUPID-Mo~\cite{Augier:2022znx}.

The prototype underwent two cryogenic tests, first in mid-2022 and then at the end of 2022, in two different configurations achieved by loading and unloading the copper spring.
We collected over 17 days of calibration across two runs with the loaded and unloaded spring. Various working points were evaluated to identify the optimal setup. As no significant differences were observed between the two configurations, this paper focuses on the results from 4 days of calibration data collected at the best working point and optimal experimental conditions during the first run with the copper spring loaded.

\section{Analysis and Results}
\subsection{Thermal properties of the tower}
\begin{figure}
    \centering
    \includegraphics[width=0.7\linewidth]{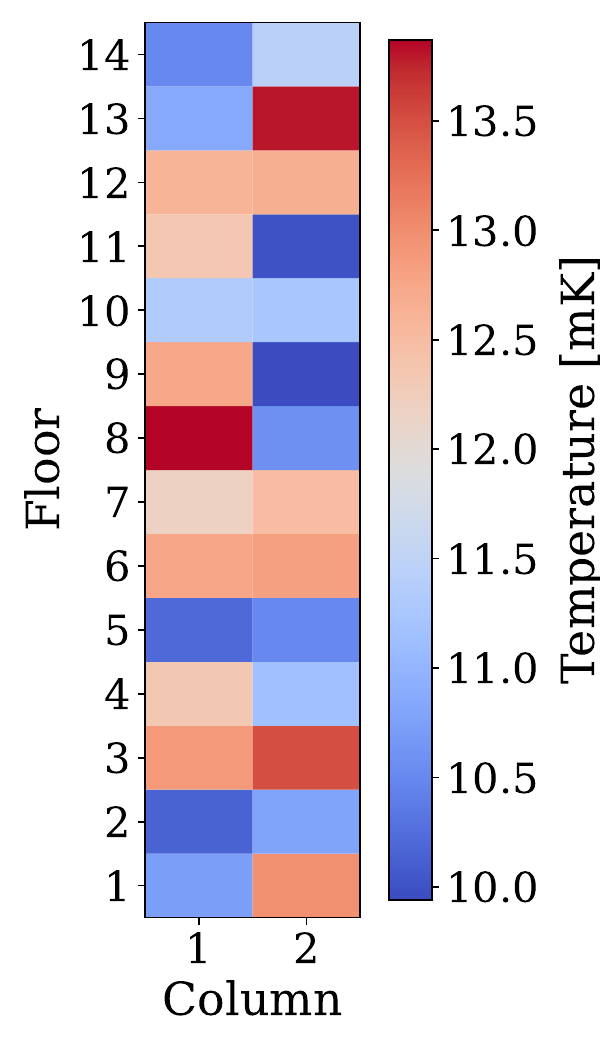}
    \caption{Base temperatures of LMO crystals along the GDPT.}
    \label{fig:baseT}
\end{figure}

\begin{figure}
    \centering
    \includegraphics[width=1\linewidth]{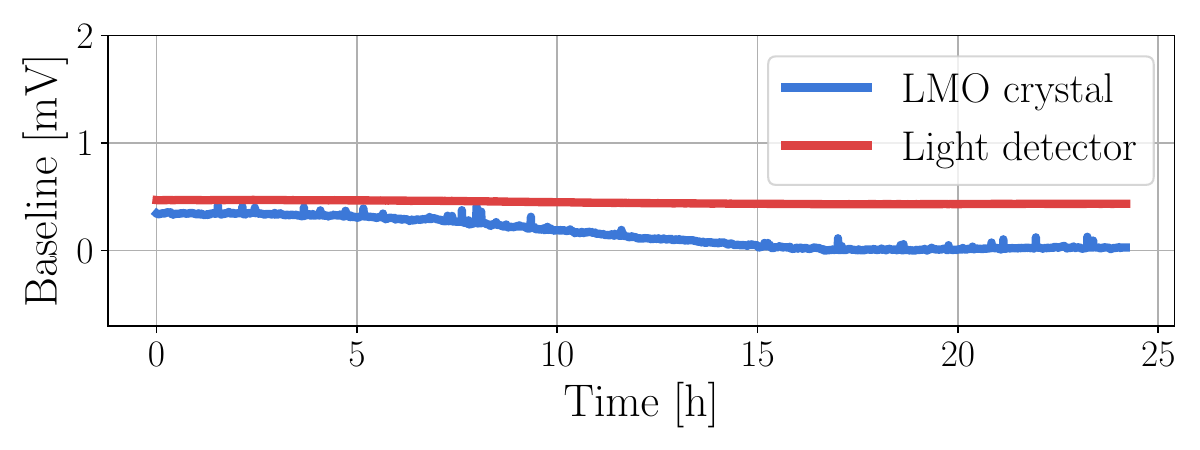}
    \caption{Voltage stream from a sample LMO-1 and LD-31 over a day of data taking. Both baselines are stable along more than 20 h, confirming the satisfying thermal properties of the GDPT. Slow variations of the baseline level were corrected during the data processing.}
    \label{fig:thermal}
\end{figure}

The GDPT successfully reached a base temperature of approximately 10\,mK, as observed with the thermometers placed along the tower.
For each LMO crystal, we measured the so-called base resistance, by biasing NTDs with a low current, $\mathcal{O}(1 \, \text{pA})$. The base resistance is the value of the NTD Ge resistance when small power 
is injected into the sensor to minimize its self-heating. Therefore, the base resistance is a good reference of the minimal temperature the NTD Ge has reached.
The resulting resistance values have been converted into temperature via the expected law $R(T)= R_{0} e^{\sqrt{T_{0}/T}}$ and considering $R_{0} = 2.4$ \si{\ohm} and $T_{0} = 3.6$ K for the specific NTD Ge batch.
In Fig.~\ref{fig:baseT}, we provide a graphical representation of the temperature uniformity along the tower. It is important to know that the observed spread in temperature is still within the accepted fluctuations of the NTD Ge characterization from one setup to another.  The resistance measured for each detector was consistent with the one detected with the thermometers, confirming the uniformity of the detectors' thermal responses along the tower.
We have implemented an active temperature stabilization system to ensure the stability of the cryostat and the tower. The remaining slow thermal drift observed in the LMO detectors, which are more massive than the LDs, as shown in Fig.~\ref{fig:thermal}, is corrected offline using thermal gain stabilization with constant-energy events~\cite{Alessandrello:1998bf,Alfonso:2018a:pulser}.
Operating the LDs at a slightly higher temperature allows to optimise their signal-to-noise ratio, as well as to obtain faster pulses~\cite{Beeman_2013} that would simplify the pile-up rejection~\cite{Chernyak:2012zz}. For this reason, we injected constant power through the corresponding heaters, until reaching the optimal base resistance of few M\si{\ohm}~\cite{Beeman_2013,Azzolini:2018tum}.

The following step consisted in choosing optimal working points for each LMO detector, by performing a scan of the voltage applied while injecting heater pulses with a reference amplitude, close to the expected amplitude of physical pulses in the MeV-scale. The operating voltage is the one which maximizes the signal-to-noise ratio, where the noise level has been evaluated as the standard deviation of the baseline of noise events.
As a result, the work resistances were in the 10--70\,M\si{\ohm} (with a median of 25 \,M\si{\ohm}) range for the LMOs and in the 4--30 \,M\si{\ohm} (with a median of 9 \,M\si{\ohm}) range for the LDs.

\subsection{LMO Performance}
The LMO bolometers achieve a signal sensitivity in the range of 15--270\,$\mu$V/MeV. 
We observed differences in the sensitivity depending on the type of glue. While NTDs glued with Araldite\textsuperscript{\textregistered} ~Rapid and UV-cured PER\-MA\-BOND\textsuperscript{\textregistered} 620 glue showed similar performance, featuring a median sensitivity of 118 and 98\,$\mu$V/MeV respectively, sensors coupled using Araldite\textsuperscript{\textregistered} ~Slow were less performing, reaching 34\,$\mu$V/MeV. This difference is probably related to a non-optimized deposited thickness for Araldite\textsuperscript{\textregistered} ~Slow, which was also responsible for a lower working resistance. However, this test confirmed that the UV-cured PER\-MA\-BOND\textsuperscript{\textregistered} 620 glue is a good alternative to Araldite\textsuperscript{\textregistered} epoxy and it matches the CUPID requirements~\cite{CUPID2024_bsl}.

The baseline resolution of LMOs is defined as the Full Width at Half Maximum (FWHM) of the Gaussian function used to fit the energy distribution of noise events. The LMO baseline resolution is characterized by a median of 3.1\,keV with a standard deviation of 2.1 keV. 
Again, we noted a slightly worse performance of Araldite\textsuperscript{\textregistered} ~Slow compared to the other glues. Indeed, we observe a median value of 3.2\,keV (1.8\,keV excluding TEST crystals) for the Araldite\textsuperscript{\textregistered} ~Rapid, 4.2\,keV for the Araldite\textsuperscript{\textregistered} ~Slow and 1.9\,keV for the UV-cured PER\-MA\-BOND\textsuperscript{\textregistered} 620 glue.

The characteristic rise time, defined as the time between 10\,\% and 90\,\% of the signal's rising edge, was 30--40\,ms. The decay time, defined as the time between 90\,\% and 30\,\% of the falling edge, was 110--120\,ms. These values are consistent with previous measurements~\cite{Alfonso2022,CUPID:2020itw,Alfonso2023}, confirming the reliability and consistency of the massive bolometers in the design of the GDPT.

\begin{figure}
    \centering
    \includegraphics[width=.98\linewidth]{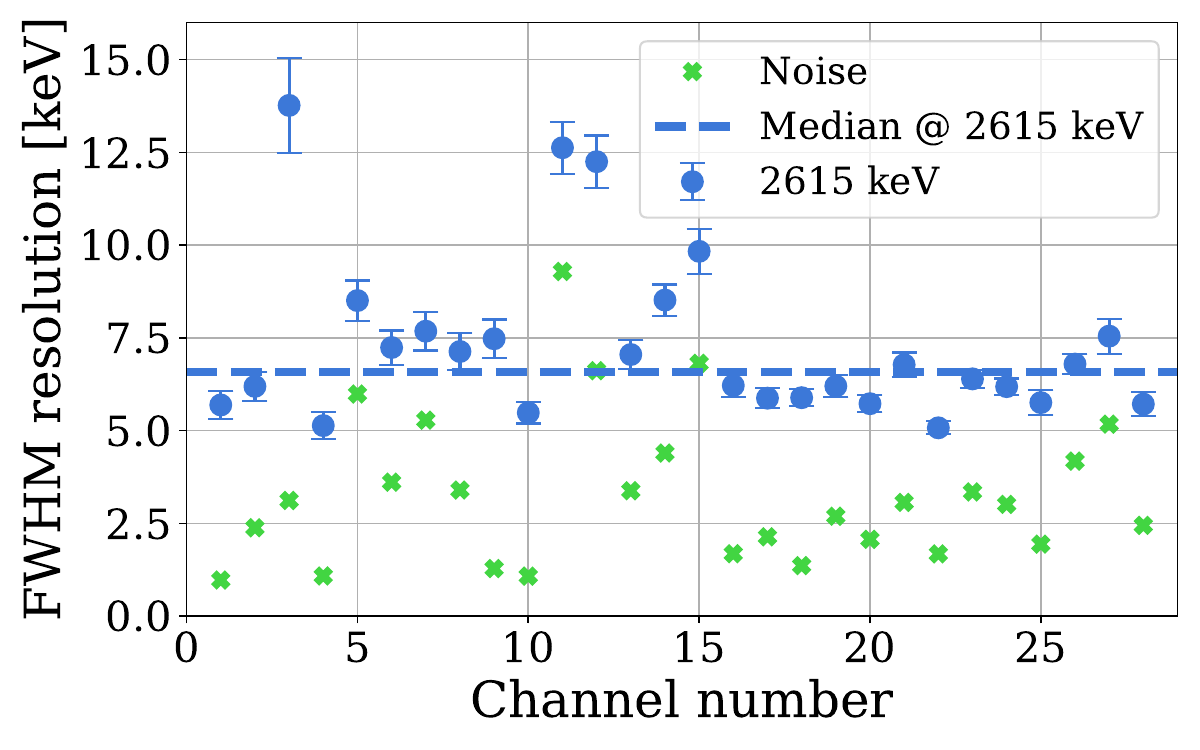}
    \caption{Energy resolution of all the LMO detectors evaluated on both noise events and events at 2615\,keV in the GDPT. The dashed blue line indicates the median energy resolution of 6.6\,keV at 2615\,keV. Excluding the TEST crystals (Channels 11 and 12), the median energy resolution would result in 6.2\,keV FWHM. We note that the statistical uncertainty for the noise resolution is hidden by the data points.}
    \label{fig:FWHM}
\end{figure}

We also evaluate the LMO energy resolution by fitting the 2615\,keV $\gamma$ peak from $^{208}$Tl with a Gaussian function on top of a linear background and quoting the resulting FWHM. 
In Fig.~\ref{fig:FWHM} we present the baseline resolution and the energy resolution at the $^{208}$Tl peak for each LMO detector.

The median energy resolution obtained was 6.6\,keV FWHM with a standard deviation of 2.2\,keV. 
We highlight that the median includes also some channels with poor performance. In particular, channel 3 (ENR-1 in Fig.~\ref{fig:detector-scheme}) was affected by additional electronics noise at 50\,Hz.
Moreover, the two Bridgman-grown crystals, channels 11 and 12 (TEST-1 and TEST-4 in Fig.~\ref{fig:detector-scheme}), exhibited slightly worse energy resolution.
However, it is important to note that these crystals were prototype crystals meant to be proof of principal for the large-scale growth of single crystals with  Bridgman method using low quality precursors. 
After these initial results, SICCAS is presently optimizing the crystal growth parameters in terms of high purity precursors, growth procedure and recovery and overall efficiency of \Mo{} isotope use. 
These growth optimizations are tested and validated in dedicated smaller fast turnaround cryogenic measurement campaigns with promising initial results.

If we exclude all the TEST crystals from the computation, the median energy resolution at 2615\,keV would be 6.2 keV FWHM with a standard deviation of 1.9 keV. This result is close to the CUPID target of 5\,keV FWHM at the $^{100}$Mo Q-value.
As expected, no further significant performance difference with respect to the crystal type was observed. Indeed, most of them were produced by the same institute following the same protocol (Section~\ref{sec:LMO}).

\subsection{LDs Performance}
\begin{figure}
    \centering
    \includegraphics[width=.98\linewidth]{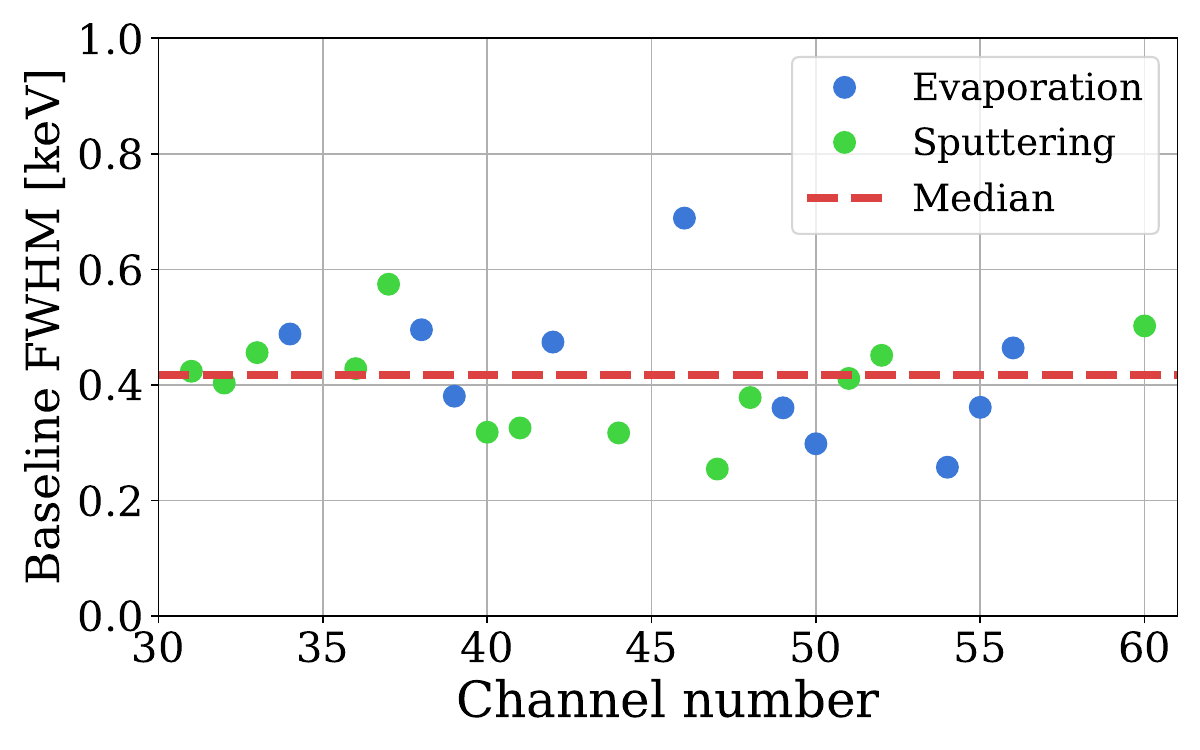}
    \caption{Baseline resolution of all LDs evaluated on noise events of the GDPT. The dashed red line indicates the median energy resolution of 0.42\,keV. Colored points refers to the type of coating. The statistical uncertainty is hidden by the data points.}
    \label{fig:FWHM_LD}
\end{figure}

The LD base and work resistances were uniform along the tower, confirming the optimal thermalization of the detectors in the GDPT structure. 
The LDs achieved a median sensitivity of 1.4\,$\mu$V/keV with a standard deviation of 0.7\,$\mu$V/keV. We also evaluated the rise time of the LDs, a key parameter for the pile-up rejection, obtaining 2--6 ms, with a median value of (3 $\pm$ 1) ms. 
The baseline resolution of LDs, evaluated as for the LMOs, varies in the range 0.3--1.1\,keV FWHM with a median value of 0.42\,keV FWHM and a standard deviation of 0.18\,keV, as shown in Fig.~\ref{fig:FWHM_LD}. These values are systematically worse than previous measurements~\cite{Alfonso2022,Alfonso_2023-pulsetube}. 
We analyzed LD noise by examining potential correlations across various noise frequencies. The analysis showed significant correlations, including at frequencies unrelated to electronics, not observed for the LMO channels. This indicates that the assembly structure of the LDs may contribute to this additional noise, particularly from the potential friction between the copper spines and the detector frames.
Therefore, ongoing tests on the optimization of the tower design are focused on reducing the Ge wafers' vibrations by improving the PTFE clamps to better constrain the Ge wafers and the overall copper structure. This is needed to avoid undesired contacts and frictional heating between the various holding elements, that we suspect to be the cause of the noise excess. 

Additionally, for the design of the next full-tower prototype of CUPID~\cite{CUPID2024_bsl}, we plan to equip LDs with aluminum electrodes in order to operate them with Neganov-Trofimov-Luke (NTL) amplification~\cite{Luke:1988xcw}. As a result of the signal amplification, the NTL gain will improve the LDs energy resolution of baseline. We expect an improvement of the energy resolution by one order of magnitude, as shown in \cite{Novati:2018zqh,CUPID2024}. Amplifying the signal will also improve the pile-up rejection \cite{Ahmine:2023xhg,Chernyak:2016aps}, which is foreseen to be a major background source for CUPID \cite{CUPID2024_bsl}. 

As with the LMOs, we tested LDs with two different anti-reflective SiO coatings. We observed no significant performance differences, both in terms of energy resolution and light collection efficiency (Section~\ref{sec:scintillation}), between LDs coated via evaporation and those coated via sputtering, indicating both methods are viable for future mass production.

\subsection{LMO Scintillation Detection}
\label{sec:scintillation}
\begin{figure}
    \centering
    \includegraphics[width=.98\linewidth]{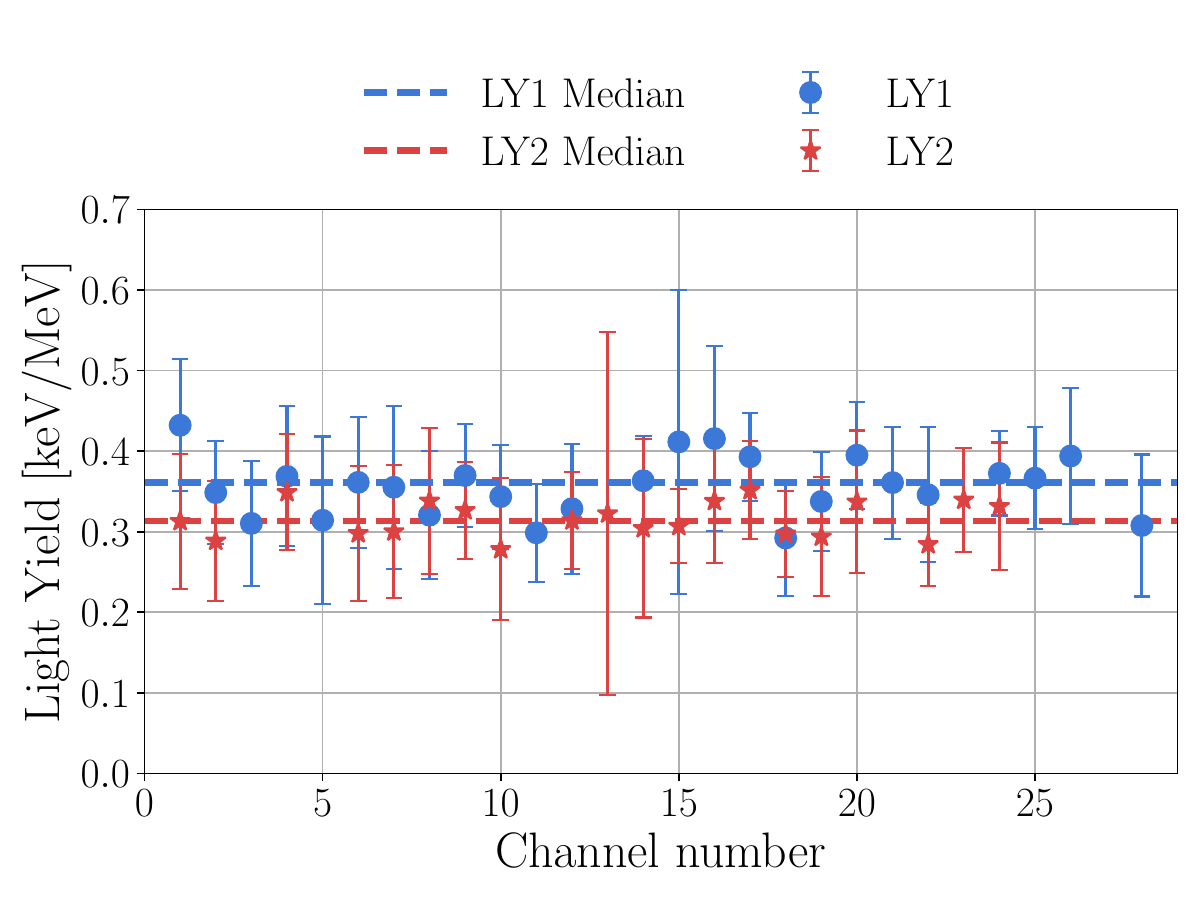}
    \caption{LY$_{\gamma / \beta}$ estimated on heat pulses in the range 2560--2670\,keV by considering the bottom LD (LY1) and the top LD (LY2). The LY1 median (0.36 keV/MeV) is reported with the blue dashed line and in red for the LY2 (0.31 keV/MeV).}
    \label{fig:ly}
\end{figure}

As part of our goals for the GDPT structure, we also measured the light yield (LY) of the LMO crystals, which is defined as the ratio between the energy reconstructed by a nearby LD and the one of the coincident heat pulse on an LMO. In particular, the LY evaluated for LMO pulses in the range 2560--2670\,keV is referred to LY$_{\gamma / \beta}$. In Fig.~\ref{fig:ly}, we present the LY$_{\gamma / \beta}$ for all LMO channels, estimated using the bottom (LY1) and top (LY2) LD. 
The median light yield of LMOs measured by the bottom (top) LDs was evaluated to be 0.36 (0.31) keV/MeV, with standard deviation of 0.04 (0.02)\,keV/MeV. 
The LY1 is systematically higher than the LY2 since the modular structure employed for the GDPT is such that the bottom LD is closer to the LMO (0.5\,mm distance) than the top one (4.0\,mm distance), increasing its geometrical light collection. It is worth noting that the measured LY$_{\gamma / \beta}$ is uniform and meets CUPID requirements also for the TEST crystals. 

\section{Conclusions}
In this work we presented the results obtained by mounting 28 Li$_2$MO$_4$ bolometers (LMO) and 30 cryogenic Ge light detectors (LDs) in an revolutionary mechanical structure, cal\-led the Gravity Design Prototype Tower (GDPT). 
We de\-mon\-stra\-ted an innovative, gravity-based approach to cryogenic calorimeter assembly, facilitating the assembly and proving a good stability and a thermal uniformity. By eliminating traditional structural supports, the GDPT reduced inert material, increasing the active detector volume with a copper/LMO ratio of less than 20\%. This improvement benefits the radioactive background contributions in CUPID through both reduced passive material and increased anti-coincidence efficiency.

Regarding thermal properties, the GDPT showed a stable base temperature at better than $\pm 0.5$\,mK stability at 10\,mK  with no systematic dependence on tower position. 
The remaining spread is consistent with the expected spread in NTD Ge sensor parameters.
The energy resolution across LMO bolometers at the $^{208}$Tl 2615\,keV $\gamma$ peak achieved a median FWHM of (6.6 $\pm$ 2.2)\,keV, (6.2 $\pm$ 1.9)\,keV by excluding the test crystals, showing no excess noise contribution and reproducing results achieved in more traditional detector holders. 
Furthermore, the GDPT showed that the (fast) UV curing  PERMABOND\textsuperscript{\textregistered} 620 glue can be a viable alternative to the Araldite\textsuperscript{\textregistered} epoxy glue used by CUORE and CUPID demonstrators.

LDs coated both via sputtering and evaporation were tested, proving consisting results.
On the other hand, the analysis of the LDs showcased an excess noise, limiting the energy resolution to 0.42\,keV FWHM (median value), despite a typical for such devices sensitivity of 1.4(7) $\mu$V/keV. This value exceeds the CUPID requirements by about a factor 2. For this reason, we are upgrading the detector design by improving the LDs anchoring in the copper frame and by removing the dual bonding of the LDs sensors in favor of a single bonding from the sensor to the CuPEN strip. Furthermore, we are upgrading the LDs technology by assisting the phonon collection with the Neganov-Trofimov-Luke effect, a voltage-driven amplification of thermal signals. A new prototype, implementing these changes, is now in construction and will take data in the first half of 2025.

The median light yield of the LMOs measured by the ``close" light detector (0.5\,mm to an LMO) was 0.36\,keV/MeV, fulfilling the CUPID’s target light yield. We note that the light yield of the test crystals produced in China was compatible with the one of the reference crystals.
The ``far" LD (4\,mm) showed anyway a remarkable light yield of 0.31 \,keV/MeV. This result highlights the feasibility of exploiting ``far" light detectors to complement the performance of the main detector. 

\section{Acknowledgments}

The CUPID collaboration thanks the directors and
staﬀ of the Laboratori Nazionali del Gran Sasso and
the technical staﬀ of our laboratories. This work
was supported by the Istituto Nazionale di Fisica Nucleare (IN\-FN); by the European Research Council (ERC) under the European Union Horizon 2020
program (H2020 /2014--2020) with the ERC Advanced Grant no. 742345 (ERC-2016-ADG, project CROSS), ERC Consolidator Grant no. 865844 (project BINGO),
and the Marie Sklo\-do\-wska-Curie Grant Agreement No. 754496; by the Italian Ministry of University and Research (MIUR) through the grant Progetti di ricerca di Rilevante Interesse Na\-zio\-na\-le (PRIN 2017, grant no.
2017FJZMCJ);  by the Italian Ministry of University and Research (MUR) th\-rough the grant ``Thin films and radioactivity mitigation to enhance superconducting quantum processors and low temperature particle detectors" (PRIN grant no. 2020\-H5\-L338); by the US National Science Foundation under Grant Nos. NSF-PHY-1401832, NSF-PHY-\-1614611, and NSF-PHY-1614611; by the Agence Nationale de la Reche\-rche,  France, through the  ANR-21-CE31-0014- CUPID-1. This material is also based upon work supported by the US Departmentof Energy (DOE) Oﬃce of Science under Contract Nos. DE-AC02-05\-CH\-11\-231 and DE-AC02-06CH\-11357; and by the DOE Oﬃce of Science, Oﬃce of Nuclear Physics under Contract Nos. DE-FG02-08ER41551, DE-SC0011091, DE-SC0012654, DE-SC0019316, DE-SC0019368, and DE-SC0020423. This work was also supported by the Russian Science Foundation under grant No. 18-12-00003.\-
This research used resources of the National Energy Research Scientific Computing Center (NERSC). This work makes use of both the DIANA data analysis and APOLLO data acquisition software packages, which were developed by the CUORICINO, CUORE, LUCIFER and CUPID-0 collaborations.

\bibliographystyle{spphys}       

\end{document}